# Experimental Investigation of the Settling Characteristics of Carbon and Metal Oxide Nanofuels

Gurjap Singh[a], Elio Lopes[b], Nicholas Hentges[a], Daniela Becker[b], Albert Ratner[a]

[a]Department of Mechanical Engineering, University of Iowa, Iowa City, IA, United States

[b]Center of Technological Sciences, Santa Catarina State University, Joinville, Brazil

**Abstract:**

Fuels dispersed with engineered nanoparticle additives, or nanofuels, are desirable for the vastly different combustion properties such as combustion rate and ignition delay they exhibit compared to base fuels. The stability of such nanofuels over time and under different particle loadings is a very important parameter to consider before they can be put into practical use. Many techniques exist today to analyze suspension stability, which have been developed to analyze water-based nanofluids. Sometimes these techniques can be expensive, and/or require specialized equipment, and/or require a method that is invasive and disturbs the suspension. Present research uses a non-contact, non-invasive, low-cost experimental setup to analyze suspension stability over long periods of time. Nanofuels made from carbon-based nanomaterials (acetylene black, multiwalled carbon nanotubes) and metal oxide nanomaterials (copper oxide, aluminum oxide) with hydrocarbon fuels (canola biodiesel, petrodiesel) have been prepared and their settling rates have been analyzed over the period of three days. It is found that metal oxides go through several metastable states as they settle. The effect of initial concentration and liquid column height is shown. It is hoped that present research showcases the positive traits of the presented technique and will spark further interest in nanofuel stability.

**Glossary:**

AB: acetylene black

BD: biodiesel

DLS: dynamic light scattering

LED: light-emitting diode

MWNT: multiwalled carbon nanotube

PD: petrodiesel

SAXS: Small-angle X-ray scattering

SSA: specific surface area

UV: ultraviolet

## 1. Introduction

The concept of nanofluids was first introduced by Choi & Eastman [1], which promised vastly improved thermal properties for a base fluid loaded with nanoparticles. Typically, these nanoparticles (which can be metallic or nonmetallic) are between 1-100 nm in size [2-3]. A small amount of Cu nanoparticles or carbon nanotubes (<1% v/v) has been reported to increase the thermal conductivity of ethylene glycol or oil by 40% and 150%, respectively [4-5].

Alumina or $Al_2O_3$, and Copper oxide or CuO nanoparticles are among the earliest investigated for their thermal enhancement, with up to 30% enhancement in thermal conductivity noted with addition of alumina nanoparticles to water [6-9]. CuO nanoparticles have also been reported to increase thermal conductivity of water up to 17% [10]. Increasingly, nanomaterials are being tested for combustion research related to enhancing the burning behavior of liquid fuel droplets, in a specific form of nanofluid which can be termed as a "nanofuel". Metal, metal oxide and carbon-based nanoparticles have been tested with an ever-increasing number of hydrocarbon-based base fuels. Researchers have investigated the effect of metal nanomaterials such as iron [11-12], boron [13], nickel [14] on liquid fuel combustion properties. The effect of metal oxides such magnesium oxide [15], alumina [16], zinc oxide [17], and cerium oxide [18] has also been examined.

It can be appreciated, however, that metals and metallic oxide nanoparticles cannot be used long term in real internal combustion engines, for the fear of metal and metallic oxide deposits on the engine walls and in the exhaust system. Carbon-based nanoparticles offer a way around that, since they are combustible themselves but can also enhance the combustion behavior of base fuel as a nanofuel. The effect of carbon nanotubes [19-20], carbon nanoparticles [21], multi-walled carbon nanotubes [21], acetylene black [22], and graphene [21] [23] on combustion properties of real-life liquid fuels has been examined in recent years.

With the rapidly increasing interest in nanofuels, and the push for nanofuels appropriate for real-world usage, the practical problem of their stability has arisen. Nanofuels are colloidal suspensions of solid material finely dispersed in a liquid fuel medium. Due to high specific surface area (SSA) and high surface energy, nanoparticles tend to form agglomerates over time, and then sediment out of the liquid fuel medium. It can be appreciated that the amount of suspended fraction inside a nanofluid strongly affects the properties that it is desirable for. From a practical standpoint, a nanofuel should stay stable over time, to which effect surfactants are sometimes added to them in small quantities. However, surfactants can themselves change fuel properties such as viscosity [24], therefore a naturally stable nanofuel is preferable to a surfactant-stabilized nanofuel.

Several techniques exist today to analyze the stability of nanofuels and nanofluids. Many researchers have tried to use specialized lab equipment for measuring suspension stability in an objective manner. A short summary of the working principles behind the various existing stability assessment techniques will be provided here.

UV-vis method works according to the Lambert-Beers law [25]:

$$A = a(\lambda)*b*c \qquad (1)$$

Where A is the measured absorbance, $a(\lambda)$ is the absorptivity coefficient dependent on wavelength $\lambda$, b is path length, and c is the analyte concentration. If I is the incident intensity of light on the sample and $I_0$ is the transmitted intensity, then $A = \log (I/ I_0)$ is the absorbance. Smaller the change in A over time, more the stability of the solution. Besides the need for specialized equipment, an obvious limitation of this technique is that all types of UV radiation are harmful to humans. Hwang *et al.* [26] have used a UV-vis spectrophotometer to analyze the suspension stability of multiwalled carbon nanotubes (MWNTs) and fullerene in oil.

Dynamic light scattering (DLS) technique measures the particle size distribution in a given sample [25]. Finding the size distribution in a sample over time gives a good estimate of the suspension stability. Notably, Anushree & Philip [25] report that this method is susceptible to the settling down of larger aggregates, and not suitable for suspensions with high concentrations because of large aggregate sizes.

Phase-contrast microscopy is a technique that relies on the phase shift in light traveling through a transparent sample, by correlating this phase shift with brightness change to obtain a magnified picture of the specimen [27]. It can be used to study the agglomeration behavior of nanofluids [25]. One of its drawbacks is that the base fluid must be optically transparent to some degree to get meaningful results. Another drawback is that agglomerates below a certain size cannot be seen using this method [25].

Zeta potential or ζ-potential represents particle surface charge, and Zeta potential values of typically 30 mV are representative of stabilized particles [28]. This technique remains an important predictor of suspension stability for many researchers, but it does have its limitations. For example, Zhu et al. [29] report during their zeta potential analysis for alumina-water nanofluids that it was not suitable for the higher-content (0.1% w/w) nanofluid and a dilute nanofluid (0.05% w/w) had to be used. Notably, nanofuels with particle loadings as high as 4% [22] are being investigated today.

Small-angle X-ray scattering (SAXS) is a technique that can characterize agglomerate sizes in a medium due to different X-ray scattering behavior of agglomerates of different sizes. Since this behavior is contingent upon the size of the agglomerate being much larger than the wavelength of the X-ray being used, sometimes only hard X-rays of small wavelength (sub-nanometer) are suitable for the application of this method in nanofluids. The equipment involved is highly sophisticated, and hard X-rays are known human carcinogens. SAXS has been used by Chen *et al*. [30] for analyzing stability of silica nanofluids.

Visual inspection, also called sedimentation method or settling bed [31-32] is a technique which has the advantage of being non-contact and non-disruptive for the nanofluid, but it is inconvenient and subjective [33]. It remains popular and widely used for the simple reason that it is easy, cost-effective, does not require special equipment, and requires only a small amount of sample. The last part is important in the context of nanofuels because many useful engineered nanomaterials (such as graphene and MWNTs) can also be very expensive.

Other specialized techniques are also being used today. Transmission electron microscopy has been used by Chiesa and Simonsen [34] to analyze the stability of alumina and oil suspensions. A new method proposed by Lee *et al.* [35] uses a hydrometer placed inside the suspension itself to monitor the suspended fraction, which is an invasive method: the presence of the hydrometer will change the settling behavior of the nanomaterials.

Anushree & Philip [24] have analyzed water-based α-$Al_2O_3$, $TiO_2$, and γ-$Al_2O_3$ nanofluids with several techniques and compared results for same nanofluids from different methods. In addition to using SAXS, UV-vis spectral analysis, zeta potential analysis, phase contrast optical microscopy and visual inspection, DLS has been used. Most importantly, all studies independently confirm the high stability of γ-$Al_2O_3$ nanofluid compared to the other ones. This shows that despite different working principles, there is correlation among these different methods.

It is the purpose of this research to present the advantages, usefulness and effectiveness of a new technique to measure suspension stability of nanofuels that is non-contact, non-invasive, requires small amounts of samples, and produces real-time and objective results that are immediately intuitive and easy to interpret.

For present research, two real-world fuels have been chosen: petrodiesel and biodiesel. Both carbon-based nanomaterials (acetylene black, MWNTs), and metal oxides ($Al_2O_3$, CuO) have been chosen to represent the most commonly used nanomaterials in nanofuels.

## 2. Methods & Materials

It has been noticed that adding nanomaterials to clear liquid fuels (such as diesel and biodiesel) changes their opacity (see Figure 1). Therefore, if a light source is shined through a freshly prepared nanofuel in which all material is well suspended, it would be mostly blocked. If the same light source is shined through the nanofuel when it has settled out at the bottom leaving only the base fuel behind, most of the light would get through. Then if a photo sensor such as a phototransistor is used to note the intensity of the light going through the sample, it will indicate the amount of suspended fraction in there. This is the working principle behind the experimental setup.

As noted before, this is a non-contact and non-invasive method, because it relies on the opacity of the nanofluid to measure the amount of suspended fraction in it, and the nanofluid itself need not be disturbed or physically tampered with. It requires low-cost and off-the-shelf electronics components such as LEDs and phototransistors and can analyze several samples simultaneously. The equipment involved works using radiation in the visible spectrum which is not harmful to humans like UV radiation or X-rays might be. The quantity of sample required is small, there is no limitation on the particle loading of the sample, and the technique itself is simple and easy to use. The method is also agnostic to composition of the nanomaterials being used: present study analyzes both metal oxide and carbon-based nanomaterials. Notably, real-time data is available as the nanofluid is settling down.

On the other hand, a notable limitation of this experimental setup is that it cannot distinguish between a large number of small particles and a small number of large agglomerates that cause the same level of opacity. This can be the case in a polydisperse system, for example. The working principle also depends on the transparency of the base fluid: an opaque base fluid is not suitable for this experimental method.

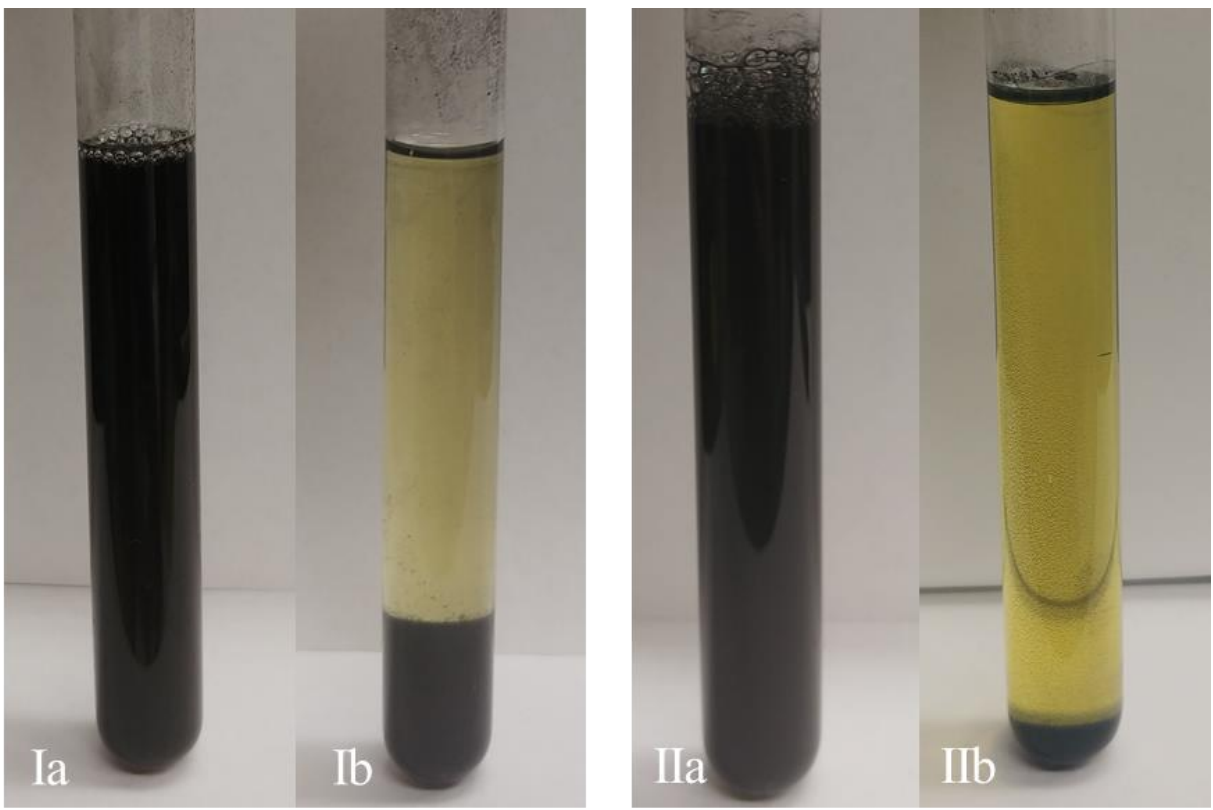

**Figure 1**. The difference in opacity between a freshly prepared and a settled-out suspension. Acetylene black in petrodiesel 1% w/w in Ia) fresh and Ib) settled out. CuO in petrodiesel 1% w/w is shown in IIa) fresh and IIb) settled out

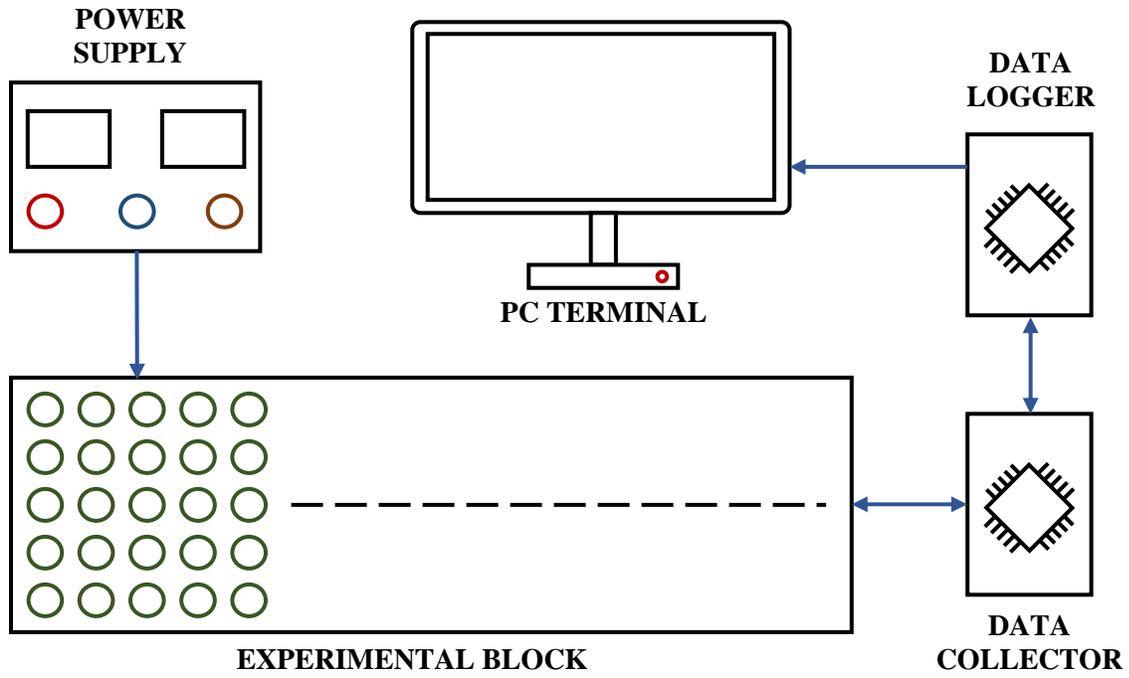

**Figure 2**. Major components of the experimental setup.

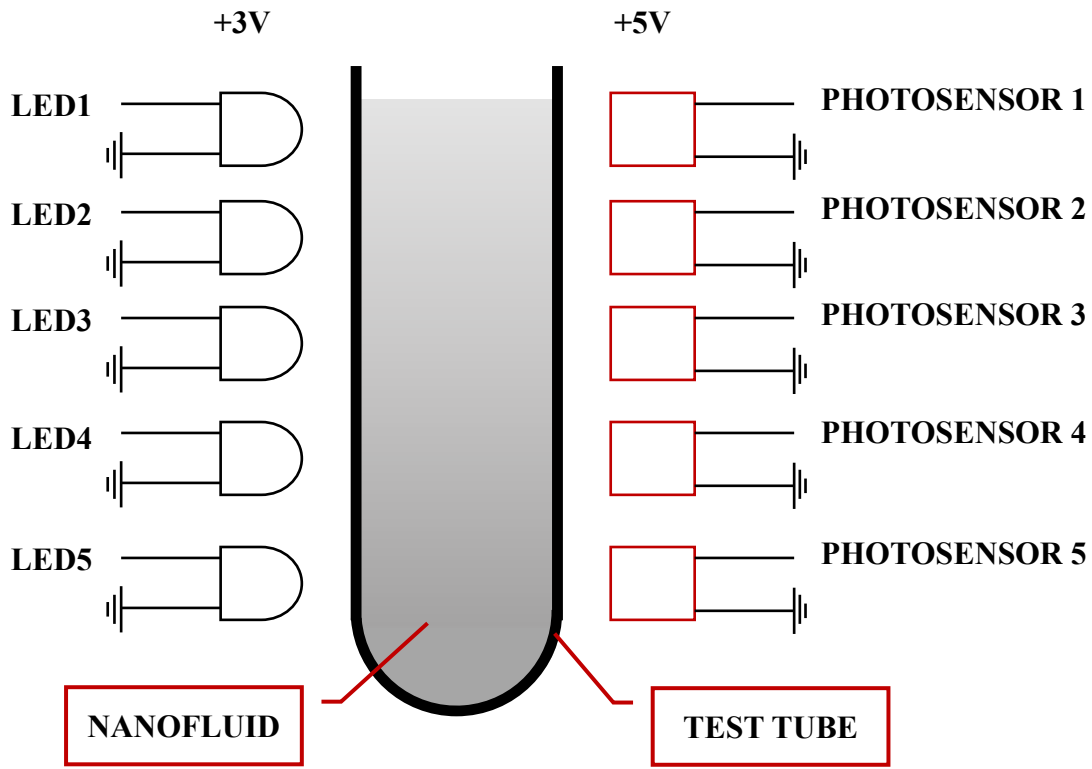

**Figure 3**. A schematic of a single experimental block showing location of LEDs and sensors

Figure 2 shows a schematic of the experimental setup, which has been detailed in the work of Singh *et al*. [33]. The experimental array consists of similarly sized rectangular spaces where a standard test tube or culture tube (Fisher Scientific) can snugly fit in. Thirty such spaces are available on the array, one space corresponds to one experimental block (see Figure 3). For one experimental block, five sensors are vertically placed at equal distance along the height of the test tube. These sensors (phototransistors, Adafruit) are placed directly opposite to bright white LEDs (Adafruit).

Acetylene black (AB) is formed by the exothermic decomposition of acetylene. It is an industrial nanomaterial known for its high electric conductivity and is used in dry cells as a charge collector [36]. It is desirable in nanofuels for its high specific surface area, combustibility, and relatively low cost.

AB used in this work was obtained from Alfa Aesar (Lot: Y04D025). It is a 100% compressed, 99.9% purity flammable solid with specific surface area 75 $m^2/g$, bulk density 170-230 g/L, and mean particle size 25-45 nm. Multiwalled Carbon Nanotubes (MWNTs) have inside diameter 3-5 nm, outside diameter 8-15 nm, length of 0.5-2 µm and specific surface area (SSA) or 233 $m^2/g$. These were obtained from Nanostructured and Amorphous Materials Inc. Copper(II) oxide or CuO was obtained from Alfa Aesar (Lot: Y19E022). It is an APS powder with SSA of 13 $m^2/g$ and particle size 30-50 nm. Aluminum oxide or alumina (Alfa Aesar, Lot Z27E054) is a 20 nm APS powder with SSA 100-200 $m^2/g$. Canola-based biodiesel was obtained from Western Dubuque Biodiesel (904 Jamesmeier Rd. PO Box 82, Farley, Iowa 52046). It is a B100 canola methylester, pale yellow fuel with 130 $^0C$ flashpoint. Petrodiesel fuel was obtained from a Phillips 66 gas station located outside Iowa City, IA.

Carbon based nanofuels were prepared by mixing the nanomaterial in 0.5%, 1.0%, and 1.5% w/w fractions with the base fuel for 24 hours on a stir plate. The metal oxide nanomaterials were prepared in the same manner but an additional step of ultrasonication for 5 minutes was added after they had been stirred for extra stability.

All nanofuels were analyzed for 3 days at room temperature. The experiments were conducted in dark and a sensor was dedicated to noting the ambient light. For all experiments ambient light levels were observed to be zero. Readings were taken at 5-minute intervals, which range from 0V (minimum) to 5V (maximum), which were normalized to 0 to 1 by dividing them with 5V. As noted before, 0 represents no light received by the sensor, which means a fully suspended nanofuel, and 1 represents maximum light received on the sensor, which means the nanofuel has settled out. In reality the signal will not reach 1 because of attenuation of the light due to a color tint to the base fuel itself, as well as refraction through the liquid medium. It will, however, reach a constant state close to 1 when nanomaterial settling is complete.

## 3. Results & Discussion

*3.1 General behavior*

Typically, a nanofuel is seen to go through three stages: settling delay, nanofuel in the settling out process, and finally the settled-out stage (Figure 4). Settling delay is noted by a "0" signal from the sensor, when the nanofuel hasn't begun to settle out yet. When it has completely settled out the signal flatlines as it reaches close to "1". Notably, the signal is never exactly "1" because the base fuel itself absorbs some of the light owing to its yellow tint. In-between these two states, the nanomaterial settles down and the signal proportionately changes. As will be explained in subsequent sections, the nanomaterial does not settle out constantly, but goes through metastable states.

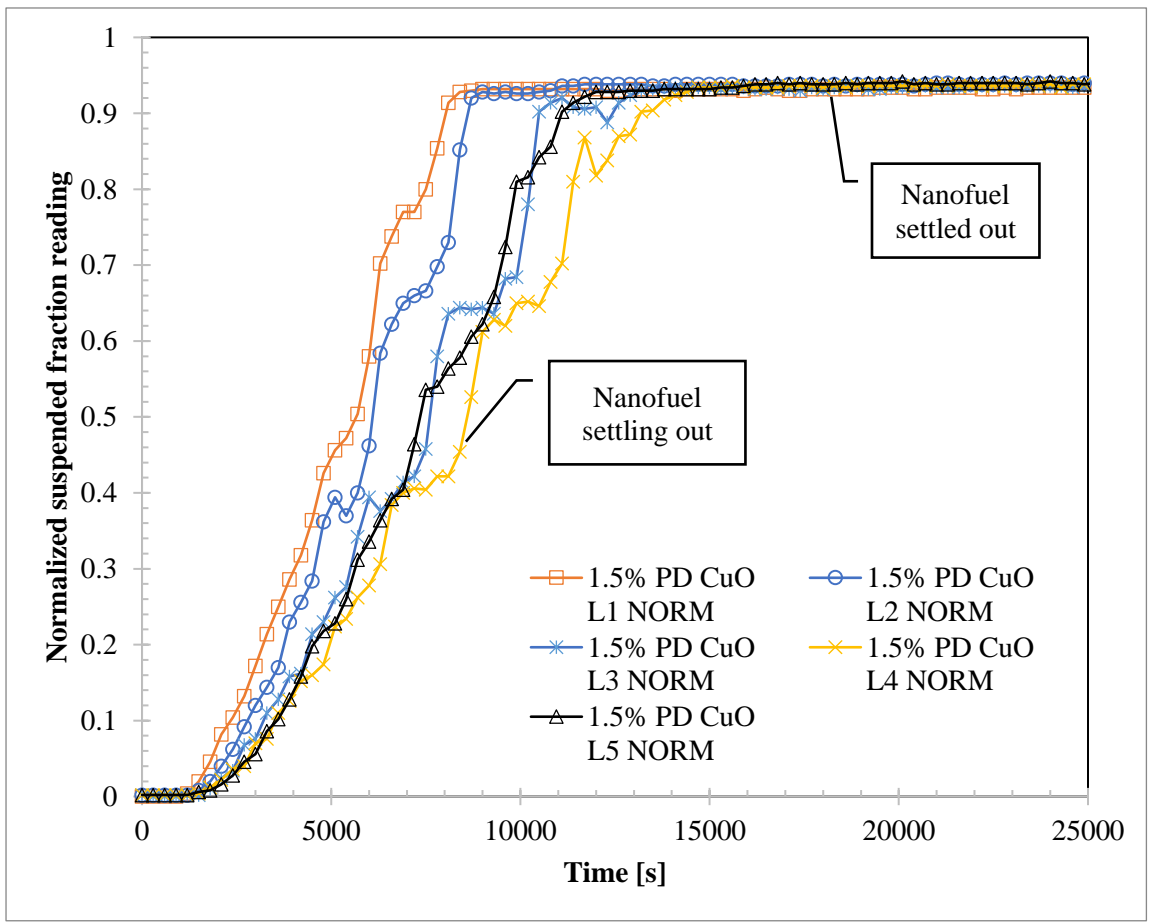

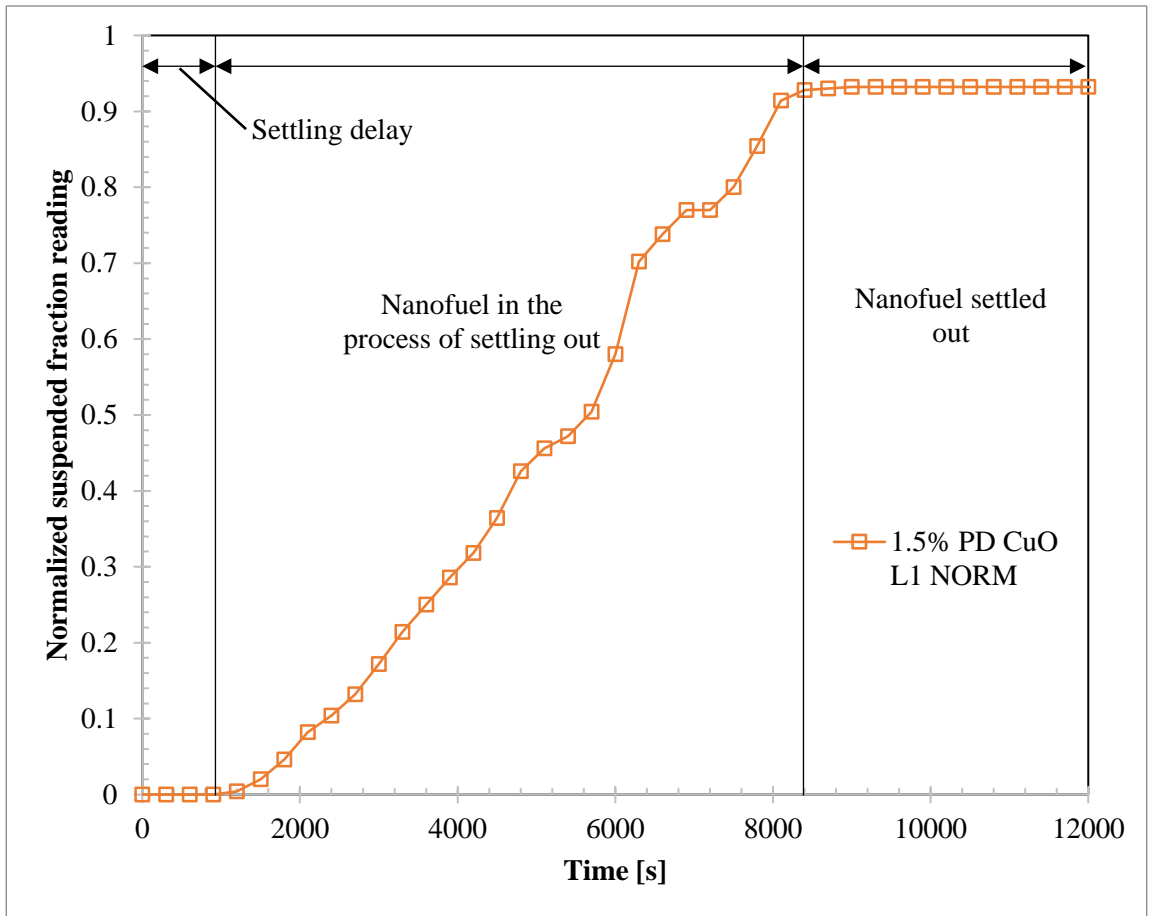

**Figure 4**. Various stages of settling in 1% CuO Petrodiesel nanofuel. (top) Normalized suspended fraction data from all five sensors, (bottom) data from the top or L1 sensor.

As can be seen from Figure 4, the representation of suspended fraction in the nanofluid is intuitive and easy to grasp, which adds to the utility of the experimental method.

*3.2 Settling time*

One of the most important parameters relevant to nanofuels is the time it takes for them to settle out completely. As is discussed later, liquid column height affects settling characteristics, therefore whichever sensor reaches the "settled out" state last is chosen to determine this. When this sensor has been at a constant state for more than 2 hours, the suspension is considered to be settled out. Figure 5 provides the settling time trends for all nanofuels analyzed during this experiment.

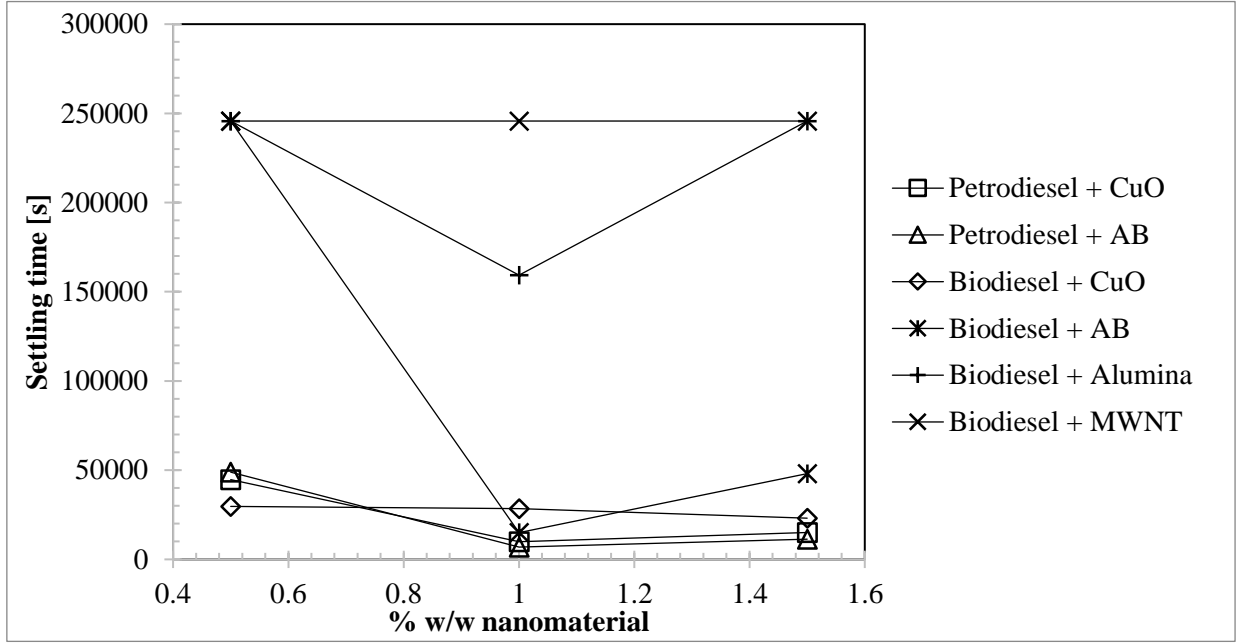

**Figure 5**. Settling time trends for different nanofuels.

It is observed that settling time is dependent on its initial particle loading. For the majority of nanofuels, 0.5% w/w particle loading is the most stable. Generally, 1% particle loading is the most unstable. Increasing particle loading after that increases the suspension stability. The MWNT nanofuel, because of the high SSA of MWNT, was the most stable nanofuels out of all that were tested. AB in biodiesel also proved to be exceptionally stable. Owing to its higher SSA, alumina made a more stable nanofuel than CuO.

*3.3 Settling trends*

The uppermost or L1 sensor sees the least effect of the liquid column, therefore it will be used to find settling trends. The settling delay and settling out processes are analyzed in this section.

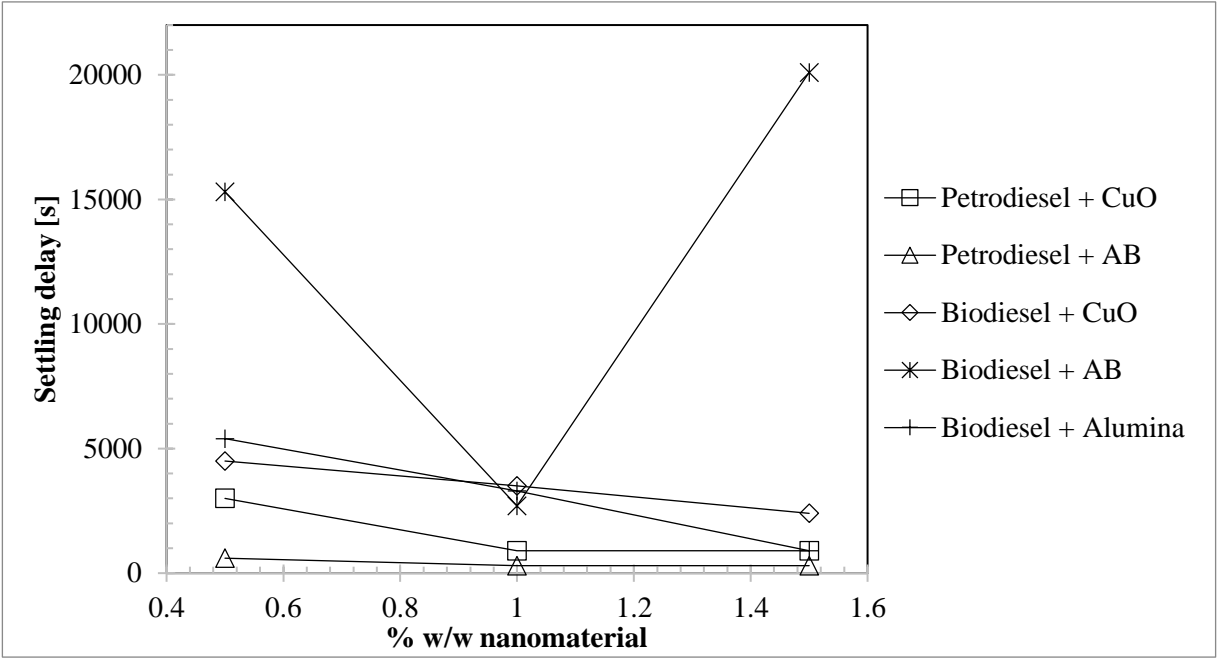

**Figure 6**. Settling delay trends for different nanofuels

Settling delays are seen to be dependent on particle loading. AB has the largest settling delay in the trends presented in Figure 6. MWNT is not represented therein as before previously noted, MWNT remained stable throughout its testing period and no settling was noted. In general, as particle loading increases the settling delay decreases. The exception to this is AB petrodiesel nanofuel, where the biggest settling delay is noted at 1.5% particle loading.

An exponential fit has been used to define the settling curves and trends. For this purpose, MATLAB® Curve Fitting Tool has been used to determine the various coefficients in a second-degree exponential relationship:

$$f = a_1 e^{tb_1} + a_2 e^{tb_2} \qquad (2)$$

Where f is suspended fraction, t is time [s], and $a_1$, $b_1$, $a_2$, $b_2$ are constants which are different for different nanofluids.

Table 1 shows the settling curve coefficients that have been determined for various nanofuels. $R^2$ shows the quality of the curve fitting, with values that are closer to 1 indicating better fit. Figure 7 and Figure 8 show some examples of the curve fitting with real data.

| Nanofuel | % w/w | $a_1$ | $b_1$ | $a_2$ | $b_2$ | $R^2$ |
|---|---|---|---|---|---|---|
| Petrodiesel + CuO | 0.5 | 0.008033 | 0.0003099 | -0.08611 | -0.0004208 | 0.9873 |
| Petrodiesel + CuO | 1 | 0.07927 | 0.0003078 | 0 | 0.0003078 | 0.9372 |
| Petrodiesel + CuO | 1.5 | 3829 | 4.02E-05 | -3830 | 4.00E-05 | 0.9918 |
| Petrodiesel + AB | 0.5 | 2.34 | -1.25E-05 | -2.444 | -3.86E-05 | 0.9935 |
| Petrodiesel + AB | 1 | 0.8642 | 4.51E-05 | 4.51E-05 | -0.09512 | 0.9894 |
| Petrodiesel + AB | 1.5 | 456.5 | 0.0002079 | -456.5 | 0.0002077 | 0.9909 |
| Biodiesel + CuO | 0.5 | 0.1051 | 9.76E-05 | 0 | 0 | 0.8576 |
| Biodiesel + CuO | 1 | 0.1343 | 7.09E-05 | 0 | 0 | 0.885 |
| Biodiesel + CuO | 1.5 | 0.4126 | 3.51E-05 | -1.044 | -0.0002405 | 0.9884 |
| Biodiesel + AB | 0.5 | -0.01145 | -4.26E-06 | 0.01234 | 2.40E-06 | 0.9741 |
| Biodiesel + AB | 1 | 0.102 | 0.02731 | -1.01E+12 | -0.6377 | 0.8475 |
| Biodiesel + AB | 1.5 | 1.49E-07 | 0.03863 | 0 | -0.1041 | 0.9877 |
| Biodiesel + Alumina | 0.5 | -0.1332 | 1.65E-05 | 0.1875 | 1.54E-05 | 0.9993 |
| Biodiesel + Alumina | 1 | 2.824 | 6.35E-05 | -2.749 | 6.39E-05 | 0.9785 |
| Biodiesel + Alumina | 1.5 | 0.2797 | 6.22E-06 | 0 | 0 | 0.9593 |
| Biodiesel + MWNT | 0.5 | 1 | 0 | 0 | 0 | 1 |
| Biodiesel + MWNT | 1 | 1 | 0 | 0 | 0 | 1 |
| Biodiesel + MWNT | 1.5 | 1 | 0 | 0 | 0 | 1 |

**Table 1**. Curve-fitting coefficients for various nanofuels

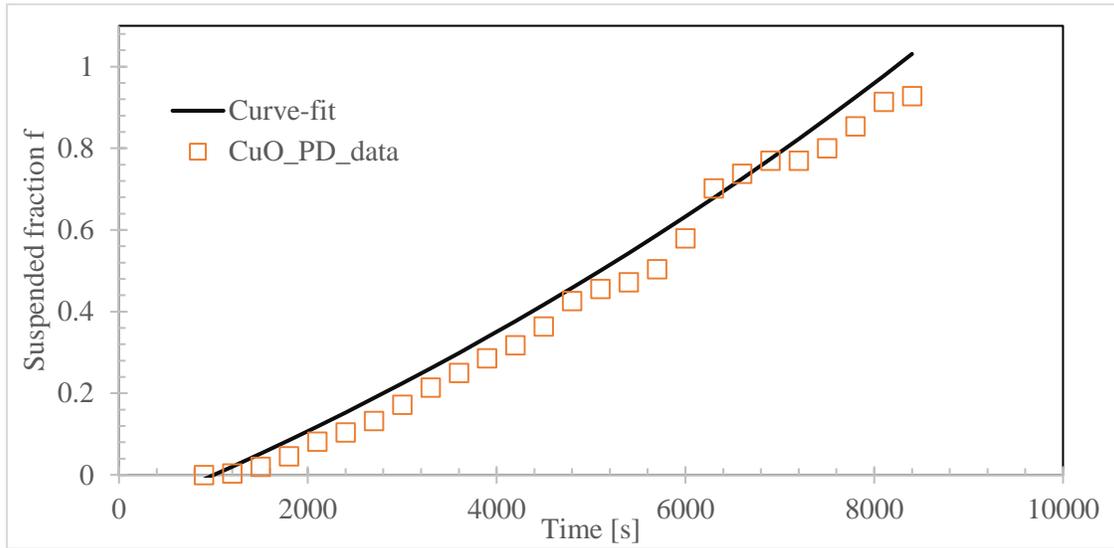

**Figure 7**. Suspended fraction for CuO petrodiesel 1.5% nanofuel trendline

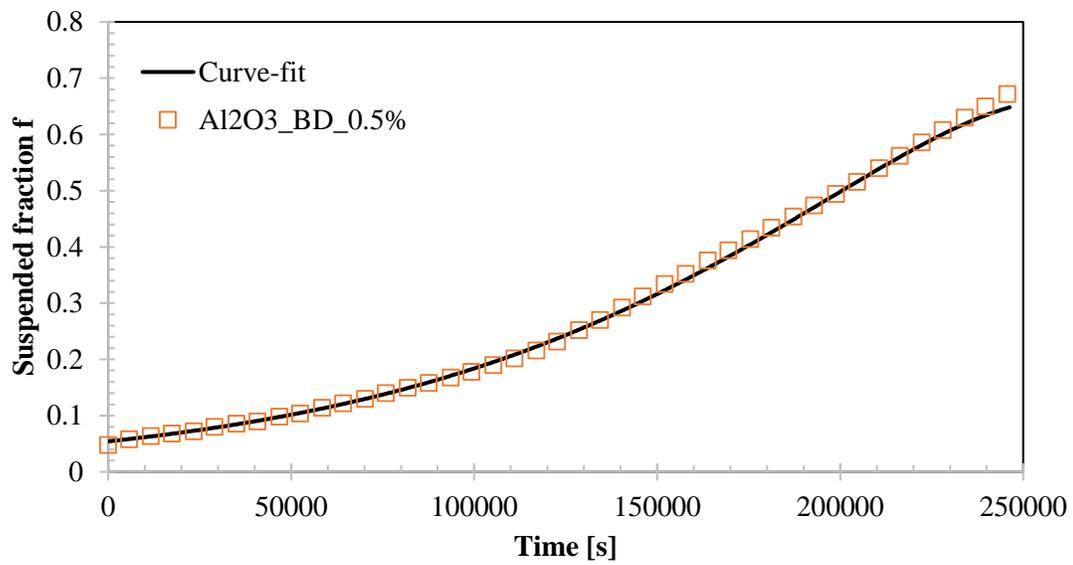

**Figure 8**. Suspended fraction for Alumina biodiesel 0.5% nanofuel trendline

*3.4 Metastable states*

It was observed that nanomaterials do not continuously settle at a constant rate, but rather seem to go through several metastable states during the process. Figure 9 shows the settling characteristics of CuO Petrodiesel 0.5% nanofuel for all its sensors. The metastable states are observable at all levels but are most pronounced at L5 because it records data at the bottom of the liquid column. IAll metastable stages are self-similar and show two characteristics: a quasi-steady state where the liquid suspension does not settle or settles very slowly, which is then followed by a very rapid settling that is characterized by a sleep increase in sensor reading. It is suggested that the quasi-steady state occurs when the nanomaterials are in the process of agglomerating to a "critical size", and when that critical size is attained the quasi-steady state ends and they settle out at a rapid rate. The process repeats with the remaining nanomaterials still suspended in the nanofuel, and keeps recurring until no more nanomaterials remain: at least 7 metastable stages can be measured for the CuO Petrodiesel 0.5% nanofuel. Note that for each stage the quasi-steady state lasts longer than the stage that came before. This is concurrent with the theory that the nanomaterials agglomerate to a critical size in this state, because as more and more nanomaterials settle out it takes more time for the remaining nanomaterials in the next metastable stage to come together and form an agglomerate that is of a sufficient size to settle down fast.

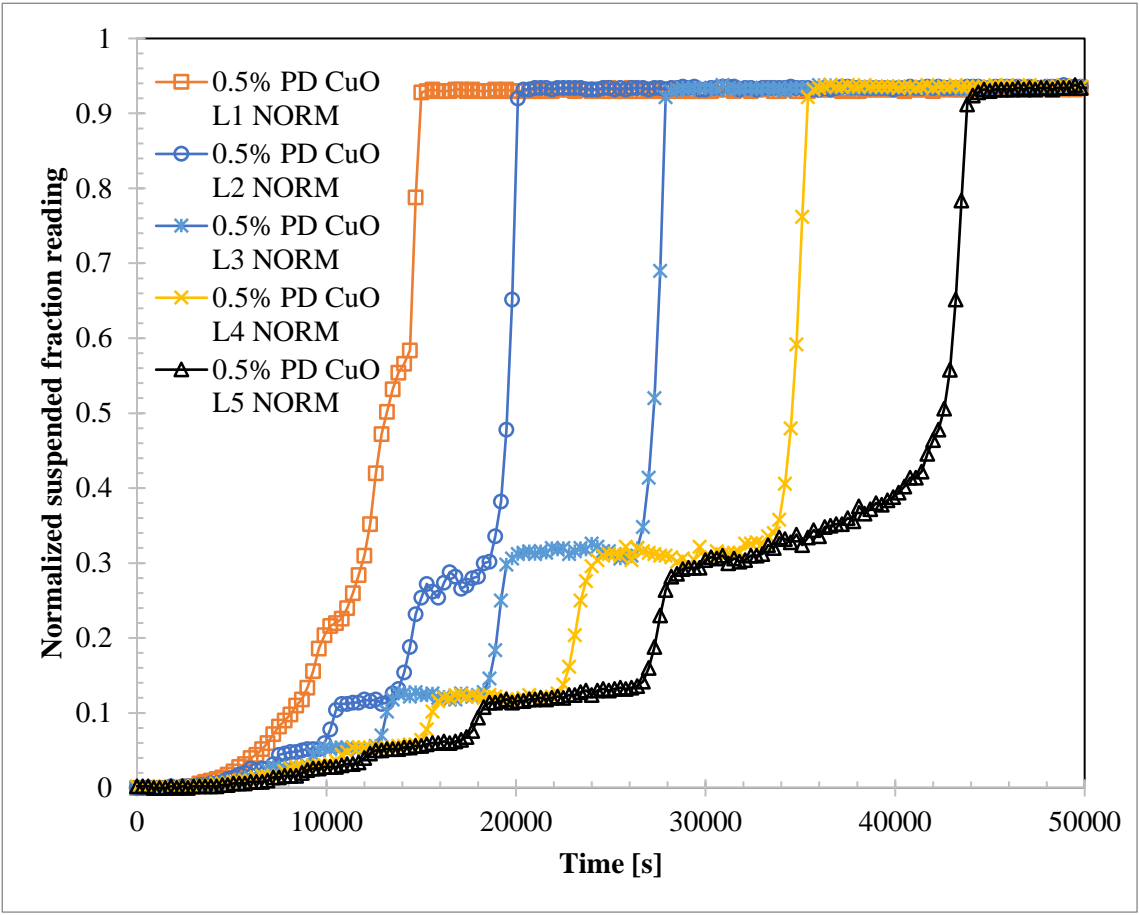

**Figure 9**. Settling characteristics of CuO Petrodiesel 0.5% nanofuel. Metastable states are more pronounced as we go down the liquid column, and L5 sensor has the most pronounced metastable states.

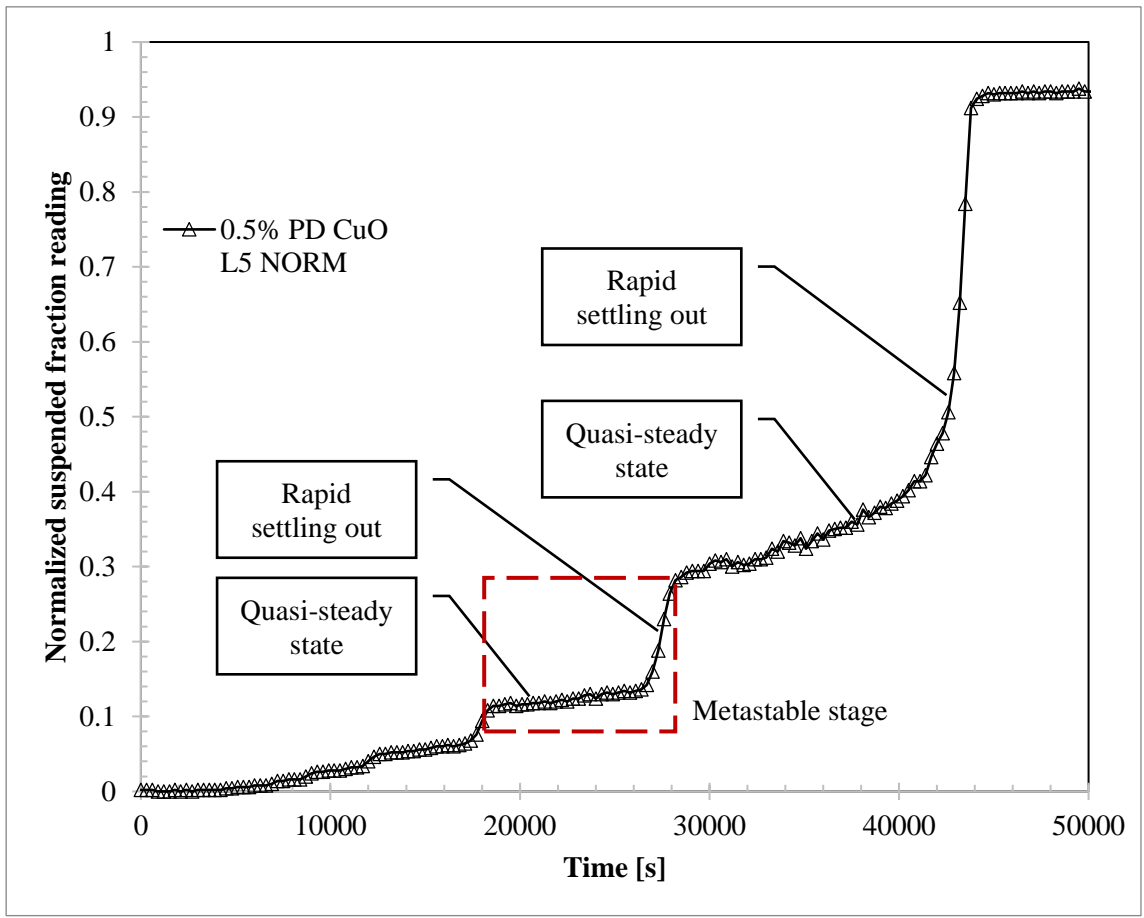

**Figure 10**. Settling characteristics of CuO Petrodiesel 0.5% nanofuel at L5 sensor. Several self-similar metastable states can be seen. Note that each lasts longer than the one before it.

## 4. Conclusions

A new method to characterize nanofuel stability has been presented in this research. This method uses off-the-shelf low-cost components to analyze the stability of a given nanofuel over time, and generates data that is simple and intuitive to interpret. It has been found that a nanofuels made from hydrocarbon-based liquid fuels and metal oxide/carbon nanoparticles go through metastable states during their settling periods. Analysis for one such case is presented. Total settling time has also been determined and has been found to be dependent on initial particle loading as well as the height of the liquid column. It is hoped that present method will supply the researchers working on nanofuels and nanofluids a convenient and simple method to analyze the stability of their working fluids.

## Acknowledgements

This research is funded, in part, by the Mid-America Transportation Center via a grant from the U.S. Department of Transportation's University Transportation Centers Program [USDOT UTC grant number for MATC: 69A3551747107], and this support is gratefully acknowledged. The authors would also like to thank Western Dubuque Biodiesel, LLC for their support. The contents reflect the views of the authors, who are responsible for the facts and the accuracy of the information presented herein and are not necessarily representative of the sponsoring agencies, corporations or persons.